# Quantum sensing with diamond NV centers under megabar pressures

Jian-Hong Dai[1]†, Yan-Xing Shang[1,2]†, Yong-Hong Yu[1,2], Yue Xu[1,2], Hui Yu[1,2], Fang Hong[1,3], Xiao-Hui Yu[1,3]*, Xin-Yu Pan[1,3,4]*, Gang-Qin Liu[1,3,4]*

[1]Beijing National Research Center for Condensed Matter Physics and Institute of Physics, Chinese Academy of Sciences, Beijing 100190, China.

[2]School of Physical Sciences, University of Chinese Academy of Sciences, Beijing 100049, China.

[3]Songshan Lake Materials Laboratory, Dongguan, Guangdong 523808, China.

[4]CAS Center of Excellence in Topological Quantum Computation, Beijing 100190, China

†These authors contributed equally to this work.

*Corresponding author.

Email: yuxh@iphy.ac.cn (X.H.Y.);

xypan@iphy.ac.cn (X.Y.P.);

gqliu@iphy.ac.cn (G.Q.L.).

**Abstract:** Megabar pressures are of crucial importance for cutting-edge studies of condensed matter physics and geophysics. With the development of diamond anvil cell, laboratory studies of high pressure have entered the megabar era for decades. However, it is still challenging to implement in-situ magnetic sensing under ultrahigh pressures. Here, we demonstrate optically detected magnetic resonance of diamond nitrogen-vacancy (NV) centers, a promising quantum sensor of strain and magnetic fields, up to 1.4 Mbar. We quantify the reduction and blueshifts of NV fluorescence under high pressures. We demonstrate coherent manipulation of NV electron spins and extend its working pressure to megabar region. These results shed new light on our understanding of diamond NV centers and will benefit quantum sensing under extreme conditions.

**Main Text:**

The development of ultrahigh-pressure techniques brings huge advances in experimental geophysics, material science, and condensed matter physics (*1*). Diamond anvil cell (DAC), which has advantages of enormous pressure range and transparent windows, is now widely used in tabletop high-pressure experiments. For geophysics, the combination of DAC and laser heating provides high-temperature and high-pressure conditions to simulate the extreme conditions of the Earth' s interior (*2*). For condensed matter physics, a number of extraordinary properties of new materials, including superconducting of hydride near room temperature (*3–5*), have been discovered under megabar pressures. However, despite these impressive advances, there have not been corresponding advances in quantitative characterizing of material properties under ultrahigh pressures, especially for magnetic, strain, and thermal properties. This is a consequence of the fact that a megabar DAC strongly constrains the sample volume (sub-nanoliter) and detection distance (> 5 mm) (*6*). As an example, Meissner effect of lanthanum hydride, an important criterion for its superconducting state (*3–5*), is still unrealized cause the samples synthesized under megabar pressure are too small (*7*).

Interestingly, a straightforward solution to the problem of high-pressure sensing is provided by diamond itself. Nitrogen-vacancy (NV) center, a point defect in the diamond lattice, exhibits superb sensitivity to local perturbations (*8–12*) and robust optical interface to sustain the high-pressure environment. A combination of laser excitation, microwave driving, and fluorescence collection could be applied to polarize, control, and read out the spin state of an NV center, providing a simple and efficient method to probe local perturbations acting on the atomic quantum sensor. This technique is called optically detected magnetic resonance (ODMR) and it sets up the foundation of color-center-based quantum technologies (*13*). The integration of ODMR and DAC techniques is straightforward, as the DAC window is transparent to the excitation laser and NV fluorescence, as shown in Fig. 1A. ODMR and quantum sensing with diamond NV centers have been demonstrated under tens of gigapascals (*10*, *14–17*). However, considering the energy level and associated optical and spin properties of NV center would be significantly modified by high pressure (*10*, *18*), whether this quantum sensor can withstand megabar pressures is still an open question. In this report, we quantify the optical and spin properties NV centers in diamond under ultrahigh pressures. We demonstrate ODMR of NV centers and extend the working pressure of diamond quantum sensing to megabar region.

We start by quantifying the optical properties of diamond NV centers under high pressures. Micron-diamonds with ensemble NV centers are placed on the DAC culet, and KBr is used as pressure-transmitting medium for its low fluorescence. A thin Pt wire embed in KBr is used to delivery microwave pulses. Figure 1B shows confocal images of the diamond culet at different pressures, measured on a home-built confocal microscopy with 532-nm laser excitation. As the pressure increases, the fluorescence from the micro-diamonds become dim but persists to pressures above 140 GPa. This is an unexpected result as a pioneering experiment has shown that the zero-phonon line (ZPL) of NV centers has a linear dependence on pressure, and a 532-nm laser cannot excite NV centers at pressures beyond 60 GPa (*10*). However, we also note a recent study reports a correction of the linear dependence of NV ZPL on pressure (*18*), which is more consistent with the current observation.

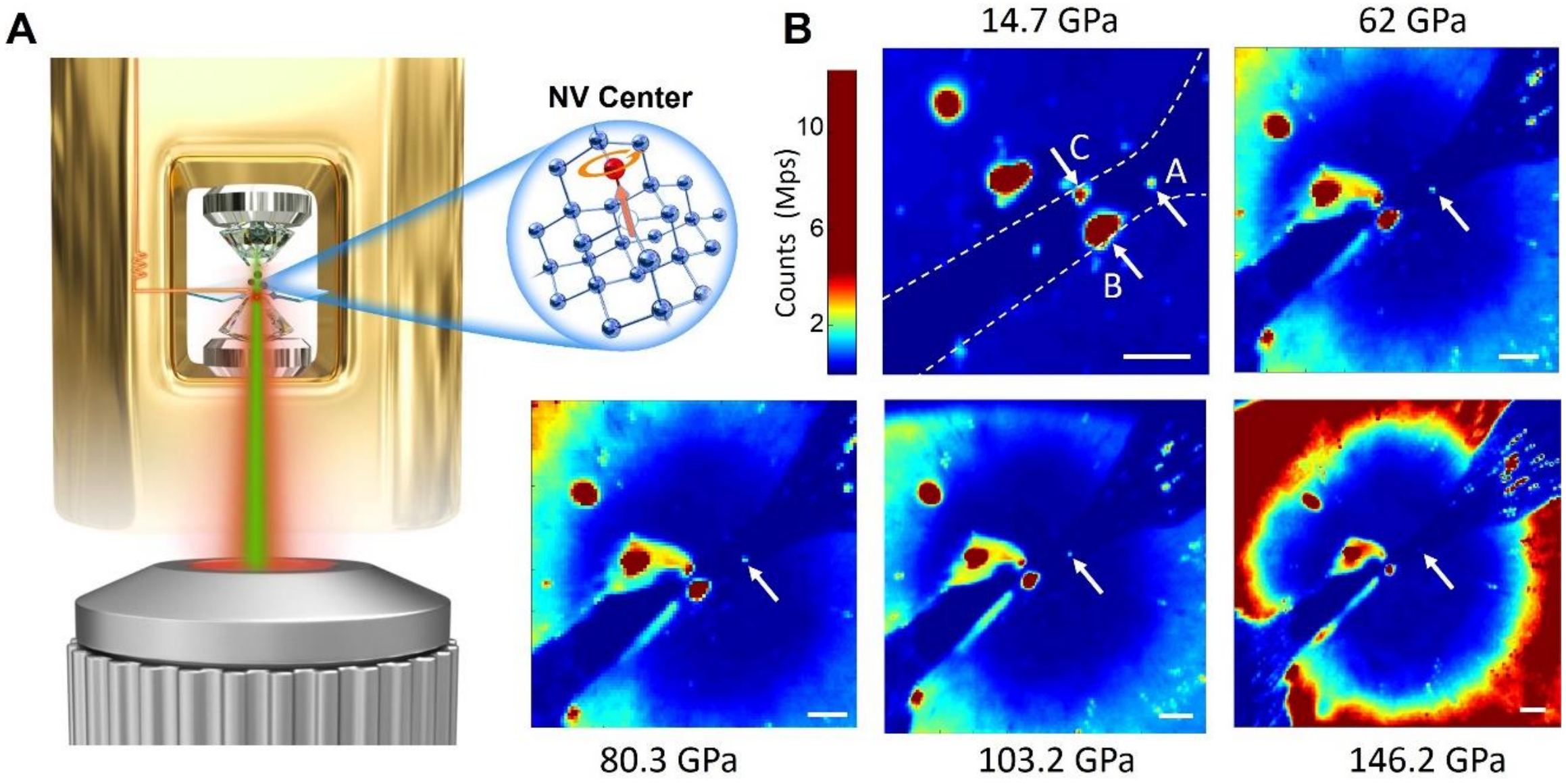

**Fig. 1. Diamond quantum sensing under megabar pressures.** (**A**) Schematic diagram of quantum sensing with nitrogen-vacancy (NV) centers inside diamond anvil cells (DAC). The precession frequency of NV electron spin carries information about its local environment. The NV spin states can be polarized with green laser pulse, manipulated with resonant microwave pulses, and read out through the spin-dependent fluorescence. (**B**) Confocal images of the DAC culet at different pressures. Diamond particles with ensemble NV centers (1-μm diameter, NV concentration ~3 ppm) are placed on the DAC culet and excited by a 532-nm laser. The white dash line denotes the boundary of a thin platinum wire, which is used to delivery microwave pulses. As pressure increases, the fluorescence of diamond NV center decreases but persist to 146.2 GPa. Scale bar is 10 μm.

To further confirm the pressure dependence of NV fluorescence at megabar region, we measure the photoluminescence (PL) spectra of microdiamonds at different pressure, as summarized in Fig. 2A. High pressures do bring a modest blueshifts to the NV fluorescence spectra, but the shift becomes slower at high pressures. The ZPL of NV centers is obscured in these spectra, as most of the emitted photons are located at the phonon side band (PSB, as depicted in Fig. 2B ) at room temperature (*19*), and inhomogeneous pressure brings extra broadenings. Figure 2C presents the fluorescence counts rate of NV centers in a microdiamond under different pressures. A dramatic drop of the NV fluorescence is observed in the first 50 GPa, and then the fluorescence signal decease slowly with pressure, consisting with the pressure dependence of PL intensity in the spectra measurement. To figure out the excitation mechanism of diamond NV centers under high pressure, we measure the count rate of NV centers in a microdiamond as a function of the excitation power, and data taken at 100 GPa is shown in Fig. 2D. The observed linear dependence indicates that a single-photon mechanism dominates the NV excitation process at 100 GPa, the same as what happens at ambient pressures. To sum up, the optical properties of diamond NV centers reveal that its energy levels are modified modestly under megabar pressures, and a 532-nm laser is sufficient to excite NV center at pressures above 100 GPa.

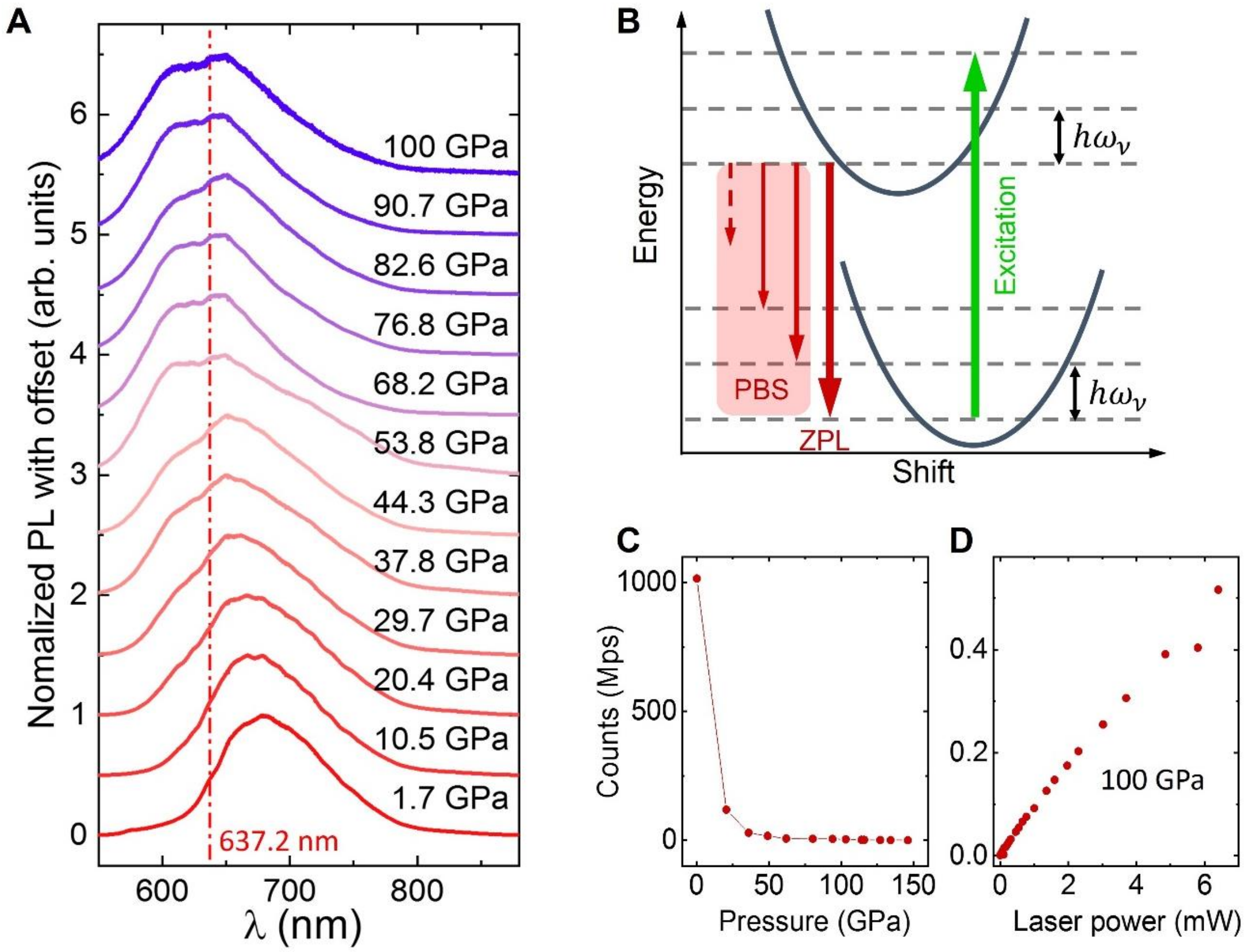

**Fig. 2. Optical properties of diamond NV centers under high pressures.** (**A**) Photoluminescence (PL) spectra of ensemble NV centers in a microdiamond under high pressures. A 532-nm laser is used as the excitation source. High pressures induce blueshifts and reduction of the signal, but the fluorescence of NV centers persists to near 100 GPa. The red dash line marks the position of NV zero-phonon line (ZPL) under ambient pressure. The ZPL peaks in the spectra are obscured by two factors: (i) most of the emitted photons are located at phonon-side band at room temperature and (ii) inhomogeneous pressure brings extra broadenings of ZPL. (**B**) Schematic drawing of the ZPL and phonon-side band (PBS) of an NV center in diamond, which will be shifted by applying pressures. (**C-D**) NV Fluorescence counts rate as a function of (**C**) the applied pressure and (**D**) the power of excitation laser. A linear dependence indicates NV centers are excited through a single-photon process. The applied pressure is 100 GPa in (**D**).

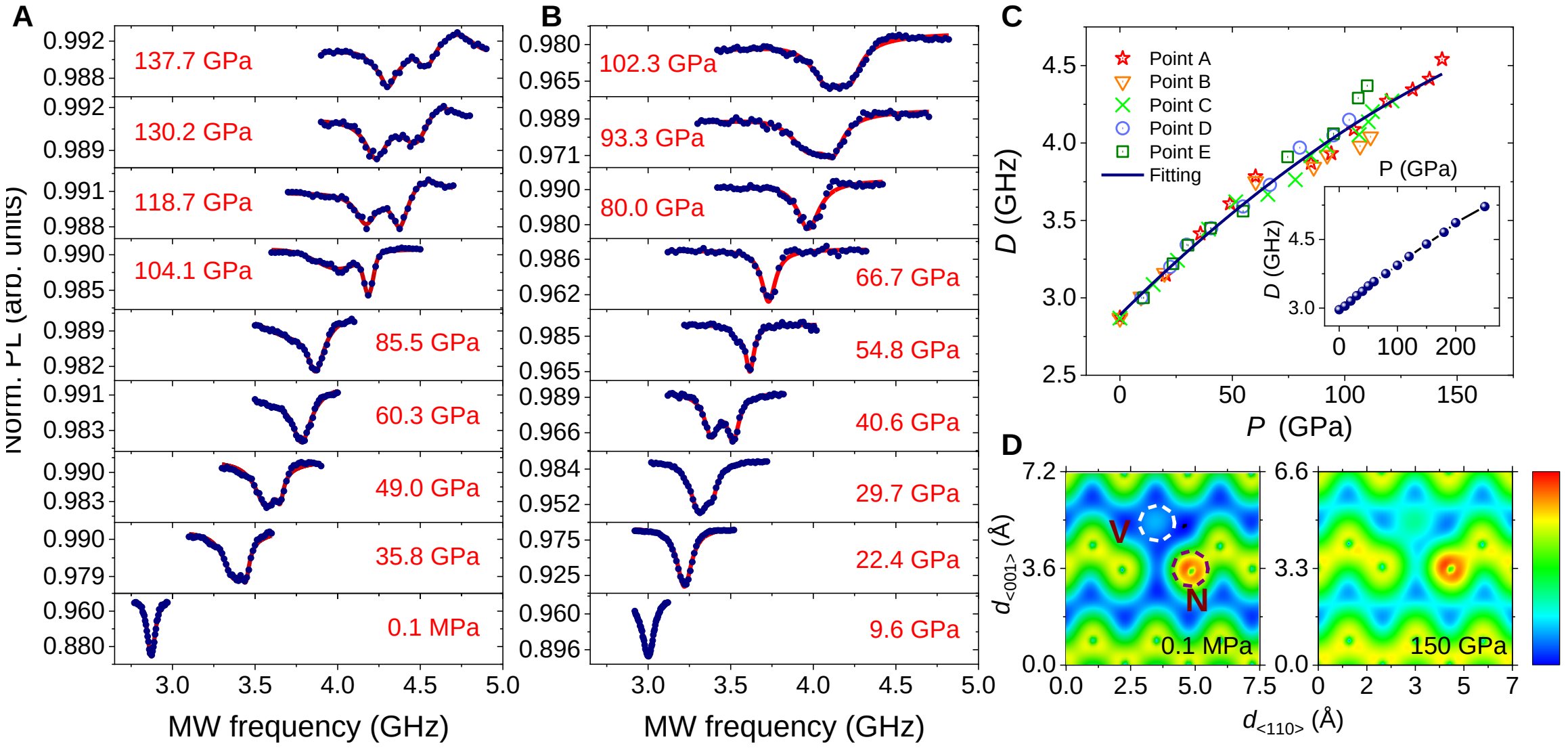

**Fig. 3 Optically detected magnetic resonance (ODMR) of diamond NV centers under megabar pressures.** (**A-B**) ODMR spectra of NV centers in two microdiamonds (point A and D) at different pressure. No external magnetic field is applied. The splitting is caused by non-axial pressure and broadening of the resonant peaks is caused by inhomogeneous distribution of pressure. (**C**) Pressure dependence of NV zero-filed splitting (*D*) for all the five measured microdiamonds, data are from two independent experiments. The solid line is a 2nd-order polynomial fit ($D = D_0 + A_1 P + A_2 P^2$) to guide the eye, with fitting parameters of $D_0 = 2.88 \pm 0.03$ GHz, $A_1 = 14.8 \pm 1.0$ MHz/GPa, and $A_2 = -27 \pm 7$ kHz/GPa$^2$. Insert: numerical results of the *D-P* relation. (**D**) Probability density of NV electron, which tends to concentrate at the vacancy and brings stronger electron spin-spin interactions at high pressures.

We then demonstrate ODMR of NV centers at megabar pressures. Figure 3A presents high-pressure ODMR spectra of NV centers in a microdiamond (point A in Fig. 1B), with no external magnetic field is applied. Before applying pressures, we observe a sharp resonance at 2.87 GHz, with a contrast of 14%, indicating the Pt wire works well in delivering MW pulses. As the pressure increases, the ODMR resonances shift to higher frequencies, a consequence of crystal field modification induced by hydrostatic pressure (*10*). Meanwhile, broadening and splitting of the resonant dips are beginning to emerge as pressure is increasing. The broadening effect is caused by the inhomogeneity of pressure among the microdiamond, and the splitting is caused by the deviation of hydrostatic condition (see discussion below). Both effects become sever as pressure increases. Nevertheless, the ODMR signal of this microdiamond persists to about 140 GPa, and its NV fluorescence gets overwhelmed by a strong background at higher pressures. Figure 3B presents high-pressure ODMR spectra of another microdiamond (point D) in an independent experiment, and similar pressure dependence is observed. Meanwhile, there is an anomalous narrowing of ODMR spectra between 40 GPa to 80 GPa for this microdiamond, which we attribute to the fluorescence quench of some groups of NV centers at high pressures. This speculation is further confirmed in the magnetic field sensing experiment (Fig. 4).

Figure 3C summaries the pressures dependence of NV center's zero-field splitting ($D$). There are totally five microdiamonds, measured in two independent experiments. Pressures below 60 GPa are calibrated with the known *D-P* relation (*10*, *14*), and pressures above 60 GPa are calibrated with the diamond Raman signal. The shift of zero-field splitting slows down at high pressures. These datasets are fitted with a second-order polynomial function, $D = D_0 + A_1 P + A_2 P^2$, which gives: $D_0 = 2.88 \pm 0.03$ GHz, $A_1 = 14.8 \pm 1.0$ MHz/GPa, and $A_2 = -27 \pm 7$ kHz/GPa$^2$. First-principal calculation is used to explore the pressure-induced modification of NV electronic structure. As pressure increases, the probability density of NV electron tends to concentrate at the vacancy, and electron spin-spin interactions becomes stronger, as depicted in Fig. 3D. The observed *D-P* relation is well reproduced with the numerical calculation (Fig. 3C insert). The nearly identical pressure dependence of all the measured NV centers provides a reliable and robust mechanism to calibrate pressure in the megabar region.

Having obtained the ODMR spectra, we now demonstrate magnetic field and strain sensing under high pressures. NV centers in a microdiamond (point D) serves as in-situ quantum sensors inside the DAC chamber. An external magnetic field is generated by a permanent magnet, and the applied pressure is about 80 GPa. In the first experiment, ODMR spectra are measured as the strength of the magnetic field is tuned, as presented in Fig. 4A. At low magnetic field, the magnetic response of NV centers is inhibited by the non-axis component of the applied pressure, and the inhomogeneous broadening further degrade the magnetic sensitivity. At relatively high magnetic fields (≥ 100 Gauss), Zeeman effect dominates again and the ODMR splitting increase with the field strength. In the second experiment, ODMR spectra are measured as the orientation of magnetic field is scanned, as shown in Fig. 4B. It is interesting to find that there are only two or four resonance dips in the spectra. Considering the geometrical configuration of possible NV structures and the large scanning angle (~ 90°) of the magnetic field, these spectra suggests that there are only two groups NV centers survived at this pressure for this microdiamond. Such a scenario could also explain the anomalous narrowing behavior of ODMR spectra of the same microdiamond (Fig. 3B). A global fitting analysis is used to track the magnetic field evolution, as depicted in Fig. 4C. The best fitting model (see Supplementary Materials for details) also gives the stress tensor of the measured microdiamond, which is $\{\sigma_{xx}, \sigma_{\mathrm{yy}}, \sigma_{\mathrm{zz}}, \sigma_{\mathrm{xy}}, \sigma_{\mathrm{xz}}, \sigma_{\mathrm{yz}}\} = \{62.8, 60.6, 75.6, -5.7, 3.8, 7.5\}$ GPa, corresponding to a hydrostatic pressure of 66.3 GPa. This value is smaller than the pressure measured on the diamond culet, indicating the existence of a ~ 10 GPa pressure gradient inside the high-pressure chamber.

To fully exploit the potential of diamond quantum sensing under high pressures, advanced quantum control techniques are needed, which relies on coherent manipulation of NV spin states (*20*). Towards this goal, we carry out Rabi oscillation of NV centers in a microdiamond (point D in Fig. 3) at about 80 GPa, as shown in Fig. 4D. The square-root power dependence of Rabi frequency is observed. Meanwhile, the Rabi oscillations present fast envelope decay, with a characteristic time of about 50 ns. This is consistent with the inhomogeneous broadening induced by pressure gradient inside the microdiamond. This Rabi signal shows that basic protocols of quantum control, like free induction decay and spin echo, could be applied to diamond NV centers under ultrahigh pressures.

We have successfully demonstrated that diamond NV center, a unique in-situ quantum sensor for high-pressure science, can be optically polarized and read out under megabar pressure. We find that maintaining of hydrostatic pressures, rather than the change of NV optical properties, is the key challenge for diamond quantum sensing at ultrahigh pressures. The observed optical and spin properties of NV centers under ultrahigh pressures deepen our understanding of spin defects in diamond, which can be extended to other color centers in diamond and silicon carbide (*21*, *22*). Considering the wide working range of diamond NV center, which covers cryogenics to 1000-kelvin temperatures (*23*), zero to tesla magnetic fields (*24*), the extension to megabar pressures marks a significant step of quantum sensing under extreme conditions. Further combination of in-situ magnetic and temperature sensing under ultrahigh pressure could enable a broad range of experiments, for example, probing Meissner effect of hydride under ultrahigh pressures, measuring the magnetic and thermal properties of solid iron at planetary core conditions (*15*, *16*, *25*).

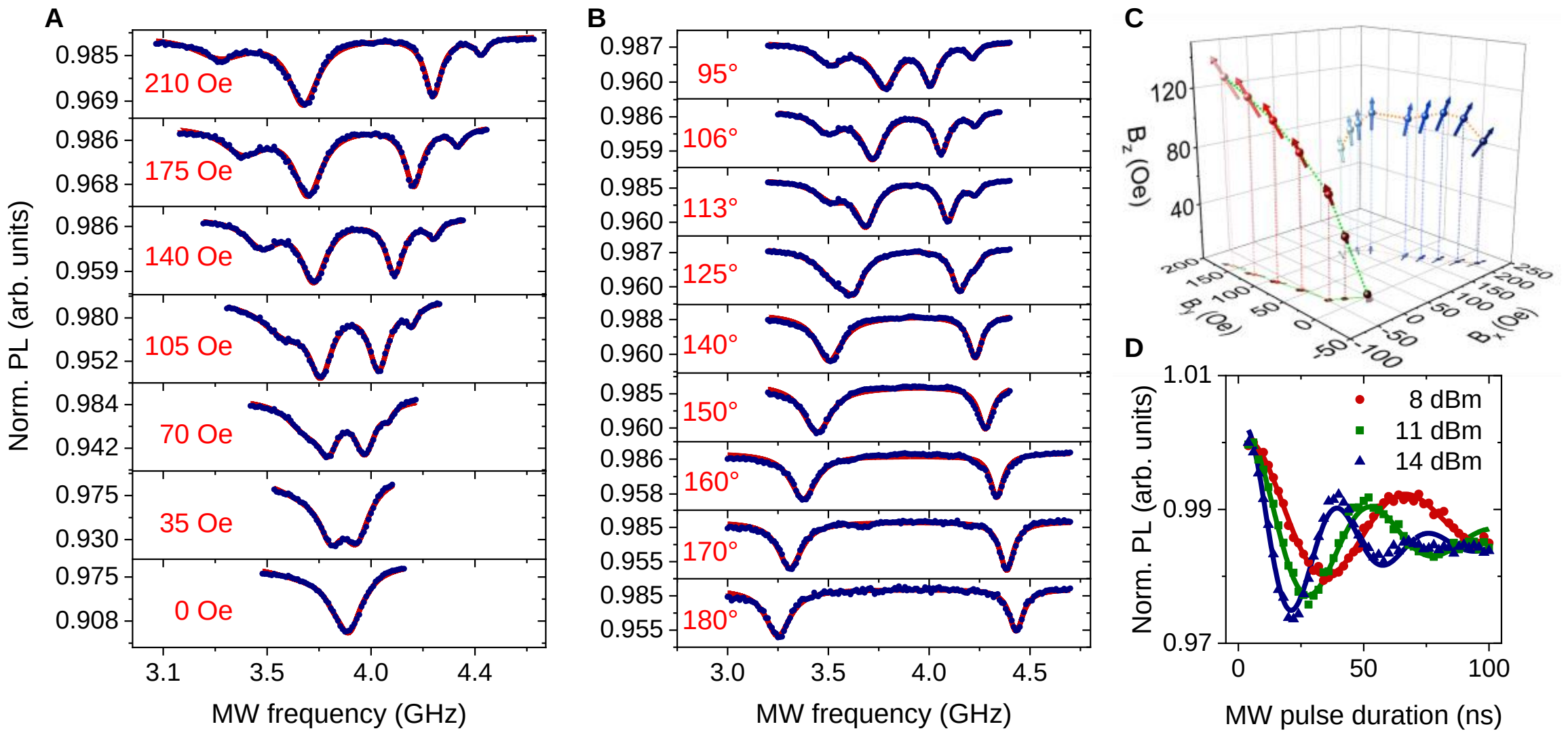

**Fig. 4 Quantum control of diamond NV centers and in-situ sensing of strain and magnetic fields.** (**A**) ODMR spectra measured under external magnetic fields of different strength. Red lines are fittings to the experimental data considering the geometrical configuration of possible NV orientations, internal stress tensor, and external magnetic fields. At low field region, the existence of non-axial pressure ($E$) decreases the magnetic field sensitivity, but this effect can be largely suppressed by a relatively large bias magnetic field ($\gamma_e B \gg E$). (**B**) ODMR spectra measured under external magnetic fields of different orientations. These spectra indicate that there are only two groups NV centers survived at this pressure. (**C**) Evolution of the magnetic field vector extracted from **A** and **B**. (**D**) Rabi oscillations of diamond NV centers with microwave driving of different powers (before feeding into the DAC). The fast decay is caused by the pressure inhomogeneity. All data shown in this figure are measured at about 80 GPa.

**Acknowledgments and Funding:** We gratefully acknowledge fruitful discussions with Miao Liu. This work was supported by the National Key Research and Development Program of China (Grant Nos. 2019YFA0308100, 2021YFA1400300, 2018YFA0305700), the National Natural Science Foundation of China (Grant Nos. 11974020, 12022509, 12074422, 11934018, T2121001), Beijing Natural Science Foundation (Grant No. Z200009), Chinese Academy of Sciences (Grant Nos. YJKYYQ20190082, XDB28000000, XDB33000000, XDB25000000, QYZDBSSWSLH013), the Youth Innovation Promotion Association of Chinese Academy of Sciences (Grant No. 202003).

**Author contributions:** G.Q.L. & X.H.Y. conceived the idea. G.Q.L., X.Y.P. & X.H.Y. supervised the project. G.Q.L., X.H.Y, Y.X.S. & J.H.D. designed the experiment. Y.X.S., X.Y.P. & G.Q.L set up the ODMR system. J.H.D. prepared the high-pressure sample. J.H.D. & Y.X.S. performed experiments. J.H.D., Y.X.S., Y.X. & G.Q.L. analyzed the data. Y.H.Y. carried out the numerical simulation. J.H.D. & G.Q. L. wrote the paper with contributions from Y.H.Y., Y.X.S., & Y.X. All authors commented on the manuscript.

**Competing interests:** Authors declare that they have no competing interests.